\documentclass[sigconf]{acmart}

\usepackage{booktabs} 
\usepackage{multirow}
\usepackage{array}		
\usepackage{tabularx}
\newcolumntype{Y}{>{\centering\arraybackslash}X}

\usepackage{graphicx}
\usepackage{subcaption}
\graphicspath{ {./figures/} }

\setcopyright{rightsretained}

\usepackage[ruled]{algorithm2e} 

\SetAlFnt{\small}
\SetAlCapFnt{\small}
\SetAlCapNameFnt{\small}
\SetAlCapHSkip{0pt}
\IncMargin{-\parindent}

\acmDOI{XXX}

\acmISBN{XXX}

\acmConference[XX]{XX}
\acmYear{2017}
\copyrightyear{2017}
\begin{document}
\title{Scalable Downward Routing for Wireless Sensor Networks and Internet of Things Actuation}

\author{Xiaoyang Zhong}
\orcid{1234}
\affiliation{%
  \institution{Indiana University Purdue University}
  \city{Indianapolis} 
  \state{IN} 
}
\email{xiaozhon@umail.iu.edu}

\author{Yao Liang}
\orcid{5678}
\affiliation{%
  \institution{Indiana University Purdue University}
  \city{Indianapolis} 
  \state{IN} 
}
\email{yaoliang@iupui.edu}

\renewcommand{\shortauthors}{ }

\begin{abstract}
In this paper, we study the downward routing for network control/actuation in large-scale and heterogeneous wireless sensor networks (WSNs) and Internet of Things (IoT). We propose the Opportunistic Source Routing (OSR), a scalable and reliable downward routing protocol for WSNs/IoT. OSR introduces opportunistic routing into traditional source routing based on the parent set of a node's upward routing in data collection, significantly addressing the drastic link dynamics in low-power and lossy WSNs. We devise a novel adaptive Bloom filter mechanism to effectively and efficiently encode a downward source-route in OSR, which enables a significant reduction of the length of source-route field in packet header.  OSR is scalable to very large-size WSN/IoT deployments, since each resource-constrained node in the network only stores the set of its direct children. The probabilistic nature of the Bloom filter passively explores opportunistic routing. Upon a delivery failure at any hop along the downward path, OSR actively performs opportunistic routing to bypass the obsolete/bad link. We demonstrate the desirable scalability of OSR against the standard RPL downward routing. We evaluate the performance of OSR via both simulations and real-world testbed experiments, in comparison with the standard RPL (both storing mode and non-storing mode), ORPL, and the representative dissemination protocol Drip. Our results show that OSR significantly outperforms RPL and ORPL in scalability and reliability. OSR also achieves significantly better energy efficiency compared to TinyRPL and Drip which are based on the same TinyOS platform as OSR implementation.

\end{abstract}

%
%




\maketitle

\section{Introduction}

Wireless sensor networks (WSNs) and Internet of Things (IoT) have been increasingly applied to various areas such as environmental monitoring, structure monitoring, smart buildings, smart cities, precision agriculture, and e-health systems. Data collection is the basic application scenario in WSNs, where sensor nodes periodically sample and transmit data packets upward to one or multiple network sink(s). On the other hand, delivering control packets downward from the sink to individual sensor/actuator nodes is also essential in many WSN/IoT application scenarios, including actuating target actuator(s), reconfiguring node parameters (e.g., sampling rate), and querying data from specific node(s). However, the WSN downward routing is significantly less studied than WSN upward routing. The major downstream protocols, such as Drip \cite{drip}, Glossy \cite{glossy}, and Opportunistic Flooding \cite{oppflooding}, are flooding based and disseminate control packets to the entire network. The lack of ability in addressing individual node(s) in those dissemination protocols makes them inefficient and impractical for network actuation in low-power and large-scale WSN/IoT deployments.

The standard RPL, the IPv6 routing protocol for low-power and lossy networks~\cite{rplrfc}, offers the capacity of downward routing, but it has been found that RPL has several significant flaws in its downward point-to-multipoint communication (e.g.,\cite{rplsurvey17, Kim:2016, orchestra, kim:2015, kim2017dtrpl, rplisit, isrplready, rplcritical, drpl}). RPL essentially suffers from the severe scalability problem for downward routing \cite{rplisit,isrplready,rplcritical,drpl}. In RPL storing mode, a node stores routing entries for all destinations in its subgraph/subtree, potentially suffering from severe scalability and reliability problem in large WSNs.  On the other hand, RPL non-storing mode uses source routing \cite{dsr} through the sink/root, which suffers from not only increased risk of packet fragmentation and thus increased battery power and network capacity consumption, but also the scalability issue of the possible length of route in a network, given constrained wireless layers, such as IEEE 802.15.4 with a maximum frame size of 127 bytes (including header) \cite{802.15.4}. Moreover, it seems that RPL (non-storing mode) might not effectively fix any unreachable failure in downward routing due to wireless link dynamics. Although recent approaches such as ORPL \cite{orpl} and CBFR \cite{cbfr} attempted to address the scalability issue of downward routing, these improvements are limited for highly resource-constrained wireless devices (see \cite{isrplready}, for example). Indeed, it is increasingly urgent to systematically study scalable, reliable and resource-efficient WSN/IoT downward routing for emerging large-scale and resource-constrained WSN/IoT system.

Source routing includes the source-route information in the packet header to route packets from the source node to destination without building and maintaining routing tables at intermediate nodes. However, a direct application of source routing (e.g., RPL non-storing mode) to WSN/IoT downward routing is problematic. First, the dynamic nature of WSNs significantly affects the reliability of the traditional source routing. The specified source-route of a control packet may be obsolete and therefore unavailable when the packet arrives at an intermediate node due to wireless link dynamics. Second, the traditional source routing does not scale well in WSNs, because physical layer protocols of WSNs are designed to have a small frame size (e.g., IEEE 802.15.4 \cite{802.15.4}) for energy efficiency. As the network diameter and hence the path length increases, containing the full source-route in a packet is inefficient and may even be infeasible. Thus, a desirable and practical source routing protocol in WSNs/IoT must simultaneously satisfy the requirements of reliability under a highly dynamic wireless communication environment and scalability for very large WSN/IoT deployments.

In this work, we present an Opportunistic Source Routing protocol, referred to as OSR, to achieve desirable scalability and reliability for heterogeneous WSN/IoT actuation. Our approach is leveraged on the recent new WSN capability of the reconstruction of upward routing paths \cite{pathfinder, Rui2013rti,Rui2015rtr}, where individual upstream data packet paths from WSN nodes to the sink can be reconstructed at the sink with a minimal overhead of path encoding piggybacking to each data packet and updated in every data collection cycle. Our designed OSR protocol introduces opportunistic routing into the source routing, which is based on the parent set \cite{eer} of a node's upward routing, to exploit alternative downward paths to address wireless link dynamics. We devise a novel adaptive Bloom filter mechanism to efficiently encode and compress the source-route path. The probabilistic nature of the Bloom filter passively enables opportunistic routing for downward packet forwarding. In addition, when a downward link between a parent node and its child node is broken, active opportunistic routing is activated to find one or more other parent(s) in the child's parent set to continue the downward forwarding. OSR only requires that each node store its direct child set rather than its entire subgraph of descendants as in RPL (storing mode) or in ORPL (compressed entire subgraph) for making downward routing decision, and therefore, OSR is extremely scalable for constrained WSN/IoT sensor/actuator nodes. The proposed OSR is general and independent of the underlying link layer.

\begin{figure}
	\includegraphics{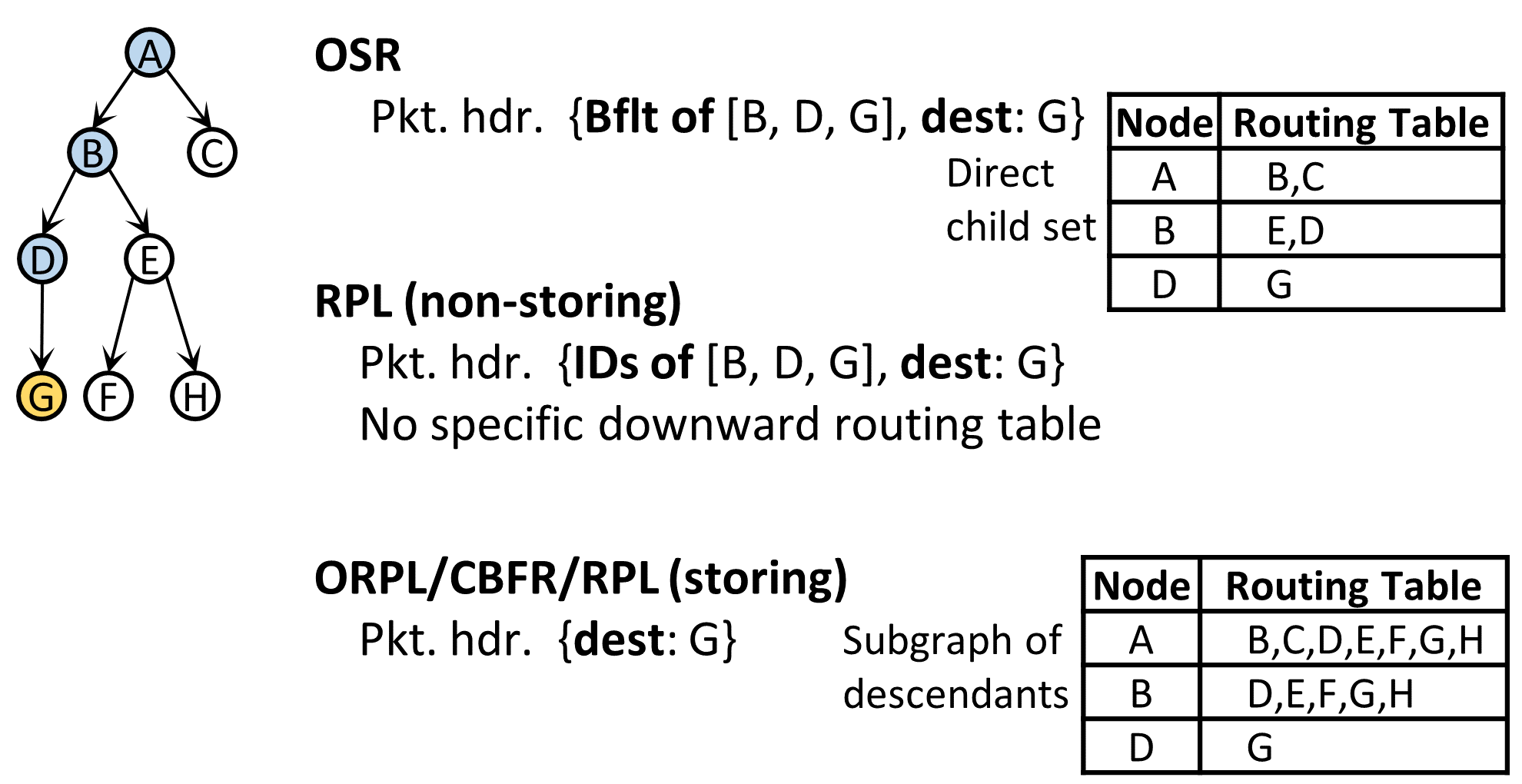}
	
	\caption{A conceptual illustration of downward packet delivery with OSR versus RPL/ORPL/CBFR.}
\label{fig1}
\end{figure}

To illustrate, Fig.~\ref{fig1} shows an example of WSN/IoT downward actuation. To deliver a packet to node $G$, OSR includes the source-route encoded using Bloom filter and the destination in the packet header; each node only stores its direct children. In contrast, RPL (non-storing mode) specifies the raw source-route and destination in the packet header, whereas RPL (storing mode)/ORPL/CBFR only specifies the destination in the packet header, with each node storing the entire subgraph of its descendants either uncompressed in RPL or compressed using Bloom filter in ORPL/CBFR, thus suffering from scalability problem.

The main contributions of this paper are: 

\begin{itemize}
	\item We propose OSR, a highly scalable and reliable downward routing approach for constrained heterogeneous WSN/IoT actuation. OSR introduces opportunistic routing into source routing to achieve reliable packet delivery in lower-power and lossy networks (LLNs). A novel adaptive Bloom filter is devised to significantly reduce the path representation overhead.  OSR enables each node to only store its direct child set as the node's downward routing table, and therefore scales well to large-size WSN/IoT deployments.
	
	\item We develop our OSR protocol, working with the popular Collection Tree Protocol (CTP) \cite{ctp}. OSR inspects CTP data packets to build the direct child set of each node. Thus, the energy overhead is negligible for building and maintaining the downward routing table. Downward routing paths are obtained based on network upward routing topology tomography.
	
	\item We evaluate our OSR protocol through both simulations and real-world testbed experiments. We show that OSR significantly outperforms RPL (both storing and non-storing modes) and ORPL on scalability and reliability.  OSR is also much more energy efficient than TinyRPL and Drip which are based on the same TinyOS platform as OSR implementation. 
\end{itemize}

The remainder of the paper is organized as follows. Section 2 presents the design of OSR. Section 3 evaluates OSR via both simulations and testbed experiments. Section 4 discusses our insights and the limitations of the current OSR implementation. Section 5 describes the related work in detail. Finally, Section 6 concludes our work and discusses the future work.

\section{OSR Design}

OSR introduces opportunistic routing into source routing, and creates an adaptive Bloom filter mechanism to encode the downward source-route path. This section presents the core mechanisms of OSR including path representation, direct child set maintenance, and opportunistic routing.

\subsection{Adaptive Bloom Filter for Path Encoding}
In traditional source routing, the entire raw routing path is included in the packet header. As network grows, this approach consumes too much overhead or may even be infeasible for large-scale WSNs. For instance, containing a source-route of 20 hops using two-byte short address in IEEE 802.15.4 takes nearly one third of the maximum link layer frame size (i.e., 127 bytes). Thus, path encoding becomes a necessity for source routing to scale in resource-restricted WSNs/IoT. OSR exploits the Bloom filter \cite{bflt2004,bflt2012} to encode the source-route path, that is, a Bloom filter representing the source-route is included in the packet header instead of the raw path.

Bloom filter \cite{bflt2012} is a space efficient probabilistic data structure that supports insertion and membership query. To insert an element into a Bloom filter of $m$ bits, $k$ independent hash functions are applied to deterministically generate $k$ hash values $h_i\in\{0,1,..., m-1\}$, and the corresponding bits are set to 1. For membership query, the element is hashed using the same set of hash functions. If all the $k$ bits are matched in the Bloom filter, the element is considered being included/matched. A membership query may result in false positives, but never in false negatives. The false positive (FP) rate of a Bloom filter can be calculated according the following equation \cite{bflt2012}:

\begin{equation}
p = \Big(1-\Big(1-\frac{1}{m}\Big)^{kn}\Big)^k,
\end{equation}
where $m$ is the length of the Bloom filter in bits, $k$ is the number of hash functions, and $n$ is the number of elements that are already encoded in the Bloom filter.

As an example of path encoding using Bloom filter, suppose in a large-scale WSN a path length $n=20$, the length of the Bloom filter $m=128$ bits, and $k=3$ hash functions are used. The resulted probability of a false positive match (i.e., false positive rate) is 5.29\%. Assuming each node address in the path occupies two bytes, using a 128-bit Bloom filter leads to 60\% space saving compared to the raw path representation mechanism, indicating the effective use of Bloom filter in source routing in WSNs/IoT.

\begin{figure}
	\includegraphics{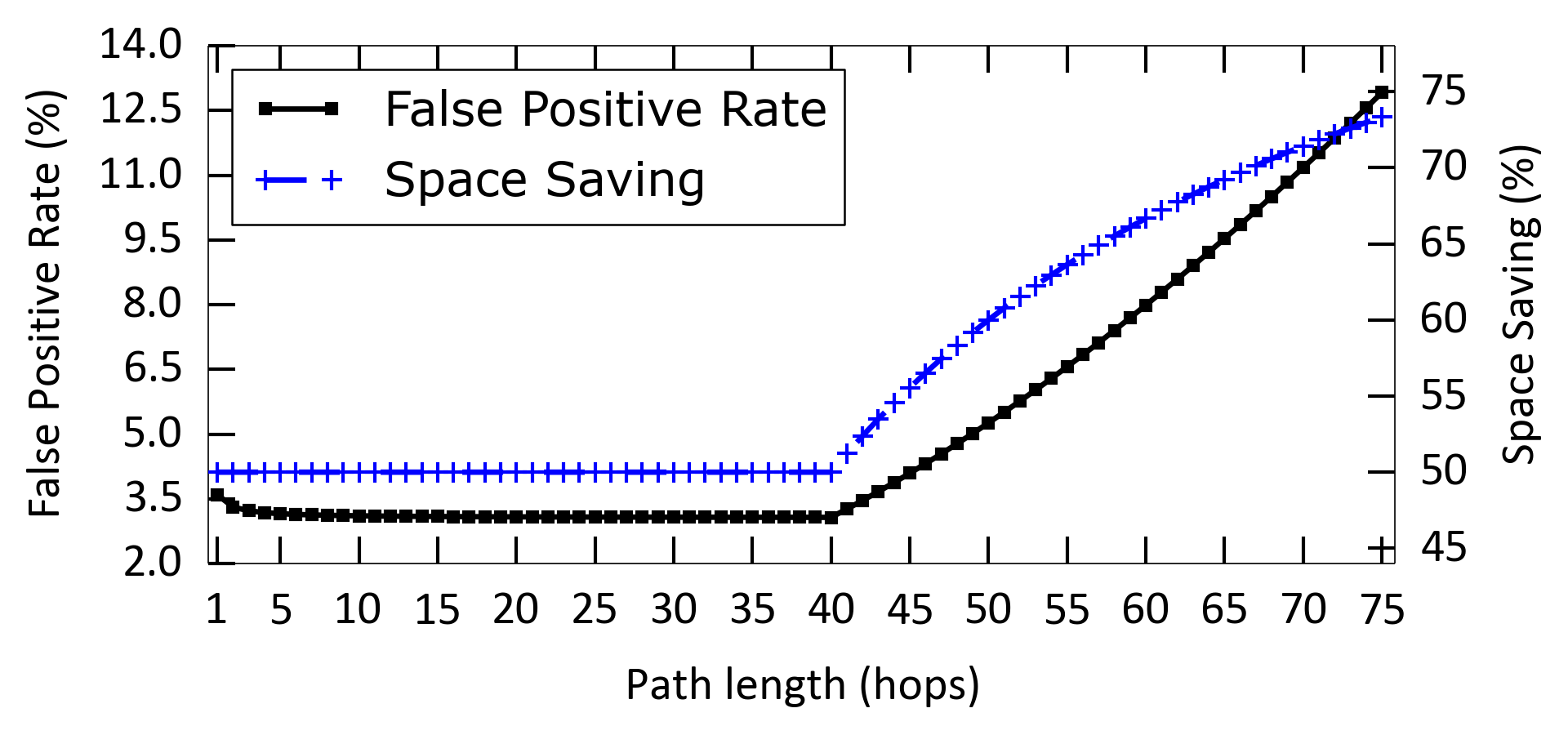}
	
	\caption{The false positive rate of adaptive Bloom filter and the corresponding space saving. Path length is $H$ hops; Bloom filter length $m$ is based on (2); the number of hash functions $k=3$; $MAX\_BFLT\_LEN=40$ bytes. }
\label{fig2}
\end{figure}

Since multi-hop WSNs usually need to be scalable in practice, using a fixed-length Bloom filter is inefficient. For instance, a too short Bloom filter would introduce high false positive rate for a long path, whereas a long Bloom filter may have more bits than a short raw path itself. We devise an adaptive path Bloom filter whose length $m$ (bits) is proportional to the hop count $H$ of the route:

\begin{equation}
m= \left\{
		\begin{array}{ll}	
			8H & H\le L\\
			8L & H>L 
		\end{array}
	\right.,
\end{equation}
where $L$ is the maximum Bloom filter length of any encoded source route in bytes. Even with a minimum node ID (i.e., address) length of two bytes, the devised Bloom filter (2) for path encoding leads to at least 50\% space saving compared to the use of raw source route in RPL (non-storing mode), indicating the potential merit of our adaptive Bloom filter. Fig.~\ref{fig2} demonstrates the analytical false positive rate based on (1) of the devised adaptive Bloom filter with $L=40$ bytes and its corresponding space saving compared to the raw path. Clearly, our devised Bloom filter mechanism scales well with respect to the raw source route length. The resulting false positive rate is lower than 3.6\% when hop counts do not exceed 40; the false positive rate actually drops as a path length increases. When a raw source route exceeds 70 hopes, our approach works fine only at a higher FP rate (<13\%). In fact, our approach would still work for any long raw path potentially of hundreds (or even thousands) of hops at somewhat degraded performance (i.e., a higher FP rate).  In contrast, traditional source routing, such as RPL non-storing mode, simply does not work for any raw path exceeding the maximum frame size of underlying link layer, which would be less than 64 hops for RPL with IPv6 address compression. If more false positives are tolerable, a shorter Bloom filter can be used to reduce packet overhead even more. In practice, a maximum Bloom filter length L (\textit{MAX\_BFLT\_LEN}) is configurable for given WSN/IoT applications. 

Due to the resource constraints in sensor/actuator nodes, hash functions employed in a Bloom filter should consume as less resource as possible. We adopt three hash functions, namely Thomas Wang's hash function \cite{hashthomas}, Bob Jenkins' hash function \cite{hashbob}, and FNV hash \cite{hashfnv} in our Bloom filter structure due to their resource efficiency \cite{cbfr}. Note that, $k$'s value can also be adaptive to the Bloom filter size and path length, as optimal $k=\frac{m}{n}\times ln2$ \cite{bflt2012}. If optimal $k$ is needed, we can further adopt SAX (Shift-and-Xor) hash to generate multiple hash values \cite{orpl, hashsax}.

When a downward packet is initialized at the sink, the source-route is encoded in a Bloom filter by ORing the Bloom filters of all the (intermediate) node addresses in the source-route. Upon reception of a downstream packet, a node checks if there exists any match of its direct child(ren) through the Bloom filter membership query of its each child, which can be done efficiently by an AND operation, as shown in Algorithm~\ref{alg:one}. If any of its direct child matches the Bloom filter, the packet is forwarded downward to the matched child node(s).

\begin{algorithm}[t]
	\SetAlgoNoLine
	\KwIn{path Bloom filter $path\_bflt$, node's hash values $h_i$, 
				$i\in\{1, ..., k\}$.}
	\KwOut{query result $ret$.}
	$ret=1$\;
	\For{$i=1$ to $k$}
		{
		\If{((($1 \ll h_i$) \& $path\_bflt$)==0) }
			{
				$ret$ = $FALSE$\; 
				
				break\;
			}
		}
	\Return{$ret$}
\caption{Bloom Filter Membership Query}
\label{alg:one}
\end{algorithm}

\subsection{Direct Child Set}
WSN/IoT nodes are usually highly resource limited. For example, a MicaZ node platform, only has 4K bytes of RAM. Even a TelosB node that is widely used in real-world WSN deployments only has 10K bytes of RAM. Existing approaches such as RPL \cite{rplrfc} (storing mode), RBD \cite{rbd}, CBFR \cite{cbfr}, and ORPL \cite{orpl} require each node to store/encode its entire subgraph of descendants for making downward routing decisions, causing the inherent scalability problem for large WSNs/IoT. In contrast, OSR only requires each node to store its one-hop direct children, referred to as the direct child set, for downward routing. Therefore, OSR is scalable on the size of the network with respect to node's memory. 

\begin{figure}
	\includegraphics{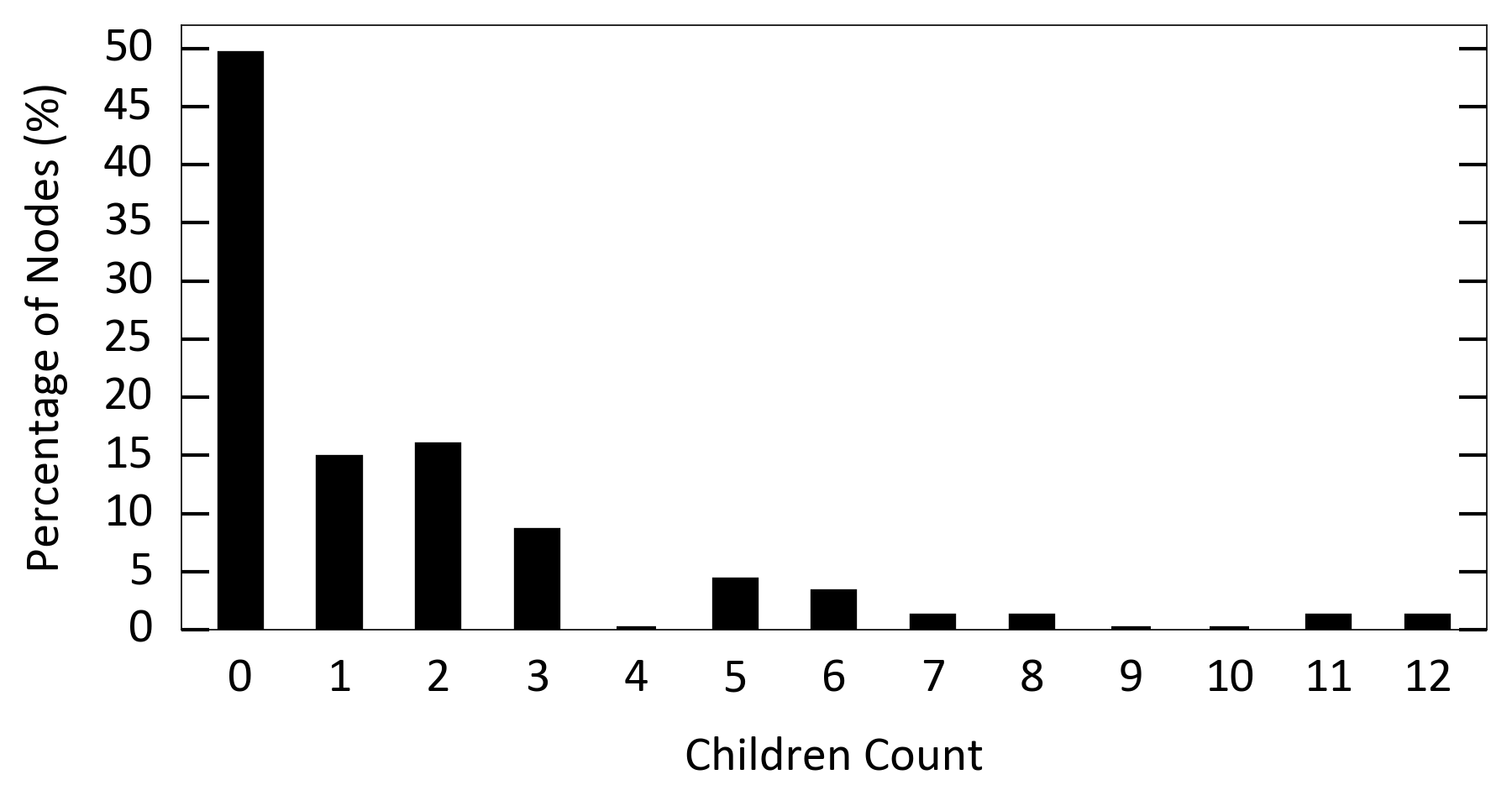}
	\caption{Node distribution on the number of direct children for the Indriya testbed with 95 nodes available. }
\label{fig:child}
\end{figure}

As an illustration, we tested a data collection WSN application based on CTP in Indriya testbed densely deployed across three floors in a school building \cite{indriya}, which contained 95 available TelosB nodes at the experiment time. The test lasted for about 6 hours. We analyzed all the (parent, child) pairs and computed the node distribution on the number of direct children they had. The statistics is shown in Fig.~\ref{fig:child}. As we can see, around 50\% of the nodes are leaf nodes, and no node has more than 12 direct children. In contrast, the subtree of an intermediate node can grow up to a size comparable to the entire network size, especially for the nodes near the sink.

OSR takes advantage of the underlying data collection routing protocol to establish the direct child set. When a node forwards an upstream packet, it inspects the packet header and adds the link layer sender's address to its direct child set. Some protocols provide easy access methods. For instance, CTP in TinyOS \cite{tinyos} offers an Intercept interface for other applications to check the contents of a forwarded CTP packet.

To accommodate wireless link dynamics and hence the coming and leaving of a direct child, each direct child is associated with a time to live (\textit{TTL}) flag. The \textit{TTL} value decreases based on a periodic timer. When a child's \textit{TTL} reaches 0, the child is removed from the direct child set. Every time a child is refreshed or added, its \textit{TTL} value is reset. With \textit{TTL}, the direct child set should be able to cover all the (parent, child) relationships in the data collection network within the time window of length \textit{TTL}. If an intermediate node is in the path Bloom filter of a downward packet but none of the node's direct child, it is highly possible that the node itself is a false positive.

\subsection{Opportunistic Routing}
In data collection WSNs/IoT, a node may have multiple candidate parents that are able to forward its data packets within a time window, forming a parent set \cite{eer} of the node. Moreover, the parent nodes belonging to a same parent set have a high probability being within the transmission range of each other. Based on these observations, OSR introduces opportunistic routing into the traditional source routing by exploring alternative routes based on node's parent set to improve the reliability of downward routing in dynamic WSNs/IoT. Note that in OSR, nodes are not aware of their parent set explicitly. Instead, a node implicitly joins a child's parent set when it adds the child into its direct child set.

The introduced OSR opportunistic routing acts in two aspects. First, the probabilistic nature of the Bloom filter of the source-route would be able to potentially, albeit in a passive way, explore the parent nodes of an in-route node not given in the source-route but opportunistically matched by false positives. This in fact provides alternative route(s), beyond the given source-route, for downward packet forwarding. During the OSR downward routing process, whenever a node has multiple matched children in the path Bloom filter, OSR transmits the packet to all the matched children by local multicast. In the case that the matched child(ren) due to the false positive(s) is/are also in the parent set of the grandchild in the downward path, the packet is then opportunistically delivered to the grandchild. Therefore OSR, to some extent, turns false positives in the Bloom filter into potential opportunities for downward packet forwarding, which can improve the reliability without any false positive recovery scheme. Second, OSR actively performs opportunistic routing by requesting the other parent nodes in the parent set of a child node to assist packet forwarding whenever a normal downstream unicast based on the source-route fails. Due to the drastic wireless link dynamics in low-power WSNs/IoT, a source-route may be obsolete when the downstream packet arrives at an intermediate node. Source routing fails if the next hop in the source-route becomes unreachable at an intermediate node. In such an event, the intermediate node will broadcast the packet to its neighborhood, hoping that one or more of its neighbors belonging to the parent set of the next-hop child node opportunistically receive(s) it. Upon reception of a broadcast packet, the node will check whether any of its direct child is in the source-route. If yes, which indicates the node likely belongs to the next-hop child's parent set, the node would forward the packet to the matched child(ren).

\begin{figure}
	\begin{subfigure}[t]{0.22\textwidth}
		\centering
		\includegraphics{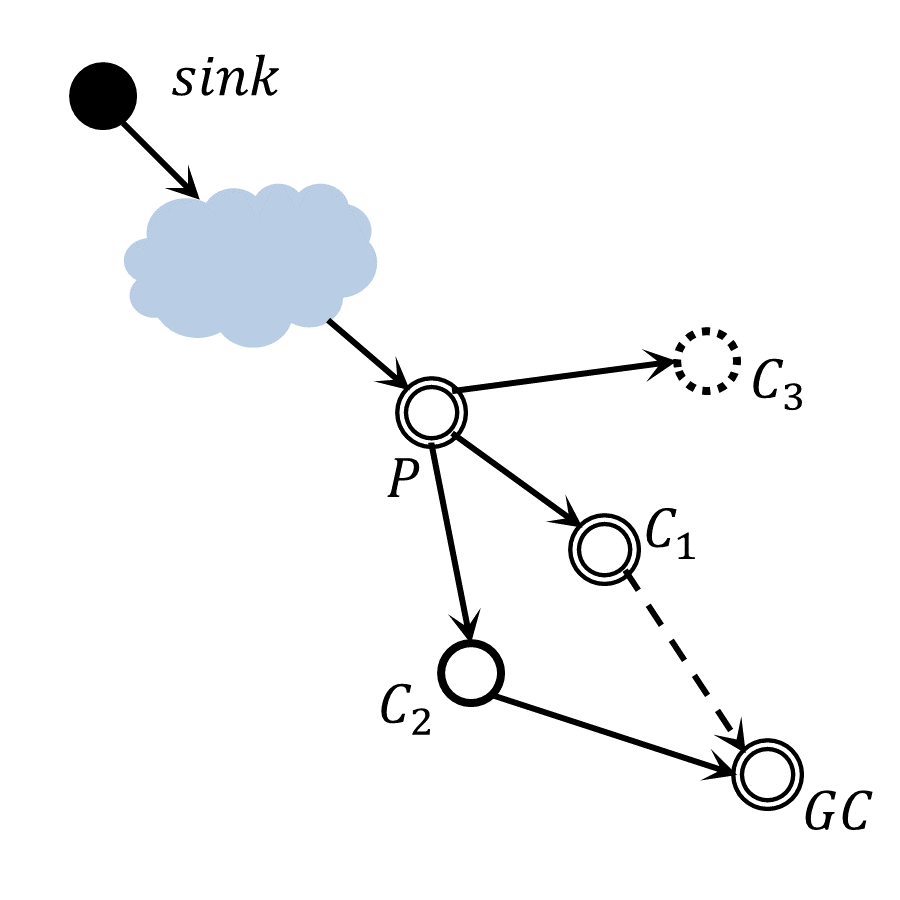}\\
		\subcaption{Passive OR}
		\label{fig:p_opp}
	\end{subfigure}%
	\begin{subfigure}[t]{0.22\textwidth}
		\centering
		\includegraphics{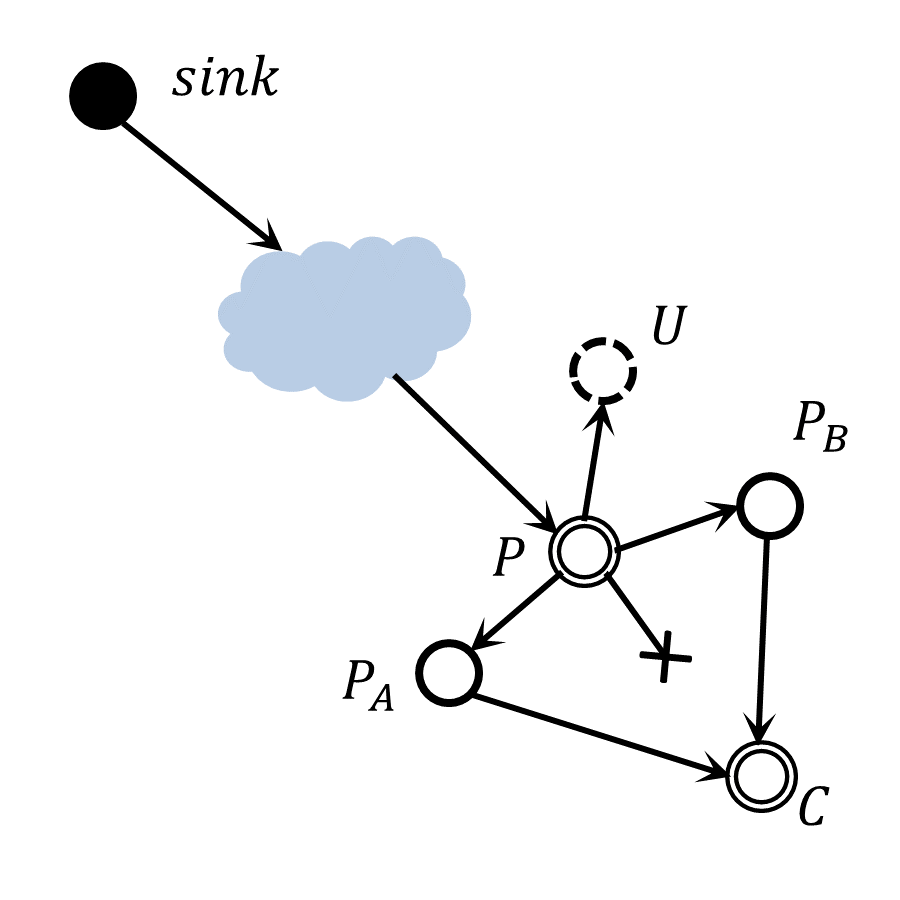}
		\subcaption{Active OR}
		\label{fig:a_opp}
	\end{subfigure}%
	\caption{Illustration of (a) passive opportunistic routing (OR) and (b) active 
			opportunistic routing (OR) in OSR. The source-route is marked as circles with double line border.}
	\label{fig:opp}
\end{figure}

Fig.~\ref{fig:opp} illustrates how opportunistic routing is conducted in OSR. The passive opportunistic routing is shown in Fig.~\ref{fig:p_opp}. The source-route specifies [$\cdots P \rightarrow C_1 \rightarrow T$]. Nodes $C_1$, $C_2$, and $C_3$ are children of node $P$ which are matched in the path Bloom filter of the downward packet. In addition to ($C_1 \rightarrow T$), node $C_2$ is also in the parent set of grandchild node $T$, hence ($C_2 \rightarrow T$) is an alternative path explored through the passive opportunistic routing. The active opportunistic routing is illustrated in Fig.~\ref{fig:a_opp}. Node $T$ is a child of node $P$ as specified in the downstream source-route. When $P$ fails to deliver the packet to $T$, it broadcasts the packet to its neighbors. Three neighbors have received the broadcast; whereas neighbor $U$ is not in the parent set of $T$ and will ignore the packet, neighbor $P_A$ and $P_B$ will forward the packet to $T$ because they are in the parent set of $T$. Thus, the obsolete link from $P$ to $T$ is successfully bypassed by the opportunistic routing activated by node $P$.

\subsection{Downward Routing Decision}
Unicast is the basic MAC layer transmission scheme used in OSR to deliver a downstream actuation packet. If any unicast fails after its maximum retransmissions, broadcast is used for active opportunistic routing. In addition, if a node has multiple direct children that are included in the source-route, it uses local multicast to deliver the packet to all the matched children for passive opportunistic routing.

OSR includes two-bit information in a downstream packet header to distinguish the three transmissions types (i.e., unicast, local multicast, or broadcast) of the packet; nodes process the received packets based on the two-bit header information accordingly. While a multicast reception would require each receiving node to check its membership to the received path Bloom filter, broadcast does not require each receiving node to check its membership. In the case of lacking the support of multicast in the MAC layer, multicast can be implemented by broadcast. OSR would benefit from multicast-supported MAC layers. Unicast reception indicates the receiver node must be included in the path Bloom filter without the check of the membership. 

We have devised the OSR algorithm (i.e., Algorithm~\ref{alg:two}) for the routing decision making process at an intermediate node. If a node receives a multicast packet and passes the membership check, it starts to check the direct child(ren); otherwise the packet is ignored. If a node receives a unicast or broadcast packet, it immediately starts to check the membership of its direct children in the path Bloom filter (\textit{path\_bflt}). If a node has multiple children included in the \textit{path\_bflt}, the packet is forwarded using local multicast. If there is only one matched child, the packet is forwarded by unicast. Any unicast failure would trigger active opportunistic routing through broadcasting. If there is no any matched child, the packet is ignored, since it is of high probability that the node itself is a false positive. In our design, each node keeps a history of recently received downstream packets to avoid duplicates and forwarding loops. Duplicate packets are ignored immediately. A time-to-live (\textit{TTL}) field (e.g., initialized as two times of the path length) is also associated with the packet to avoid infinite forwarding loops.

\begin{algorithm}[t]
	\SetAlgoNoLine
	$path\_bflt$:  the Bloom filter contains the IDs along the downstream path.\\
	$matched\_count$:  the number of matched children in $path\_bflt$.\\
	$tx\_typet$:  the transmission type of the recieved downward packet.\\
	
	\If{packet is duplicate}
		{\Return
			}

	\If {$tx\_type$ is Multicast) }
		{
			\If{local ID is NOT included in $path\_bflt$}
			{
				Ignore the packet and return
			}
		}
	Check children for match\\
	\uIf{($matched\_count > 1$)}
		{	Multicast the packet
			} 		
	\uElseIf{$(matched\_count > 0$)}
		 {
			Unicast the packet\\
			\If{unicast fails}
			{
				\If{$tx\_type$ is not Broadcast}
				{	
					/*Opportunistic routing*/\\
					Broadcast the packet
				}
			}
				
		}
	\Else{
		/*no matched children, ignore the packet*/  
	}
	\caption{Downward OSR}
	\label{alg:two}
\end{algorithm}

\section{Evaluation}
We implemented OSR in TinyOS 2.1.2, integrating CTP as the underlying data collection protocol, and performed a series of simulations and real-world WSN testbed experiments to evaluate and compare it against existing protocols. After a few initial data collection cycles, the sink starts to issue actuation commands to each individual node. We plan to make our OSR implementation, including all the test applications, publicly available.

\subsection{Methodology and Setup}
OSR is evaluated against RPL (both storing and non-storing modes), ORPL\footnote{https://github.com/simonduq/orpl}, and Drip. Drip is a TinyOS implementation of the representative dissemination protocol Trickle \cite{trickle}. We consider both ContikiRPL \cite{contikirpl} and TinyRPL \cite{tinyrpl}, the two most widely used open-sourced RPL implementations. ContikiRPL supports both storing mode and non-storing mode, whereas TinyRPL only supports storing mode. ORPL is implemented based on ContikiRPL storing mode \cite{orpl}. The test application of RPL is written based on the examples that come with the RPL and ORPL implementations.  OSR's direct child set was configured to 20 with the child \textit{TTL} value initialized to 4, which was updated every collection cycle.

We consider the following key performance metrics. First, scalability, indicated by the protocol's performance as downward path length increases. Second, the network downward Packet Delivery Ratio (PDR), defined as the ratio between the number of actuation packets received by the target nodes and the total number of packets sent by the sink. Then, the Duty Cycle (DC), the portion of time when the radio is on in low power MAC, as the measurement of energy efficiency with the same implementation platform. For all the performance results the sink's transmissions are not included, as the sink/root is considered to have unlimited power supply. 

\subsection{Evaluation in Cooja}
We first conducted simulations in Cooja \cite{cooja} using TelosB platform to evaluate the scalability of OSR against RPL and ORPL. 

\begin{figure}
	\includegraphics{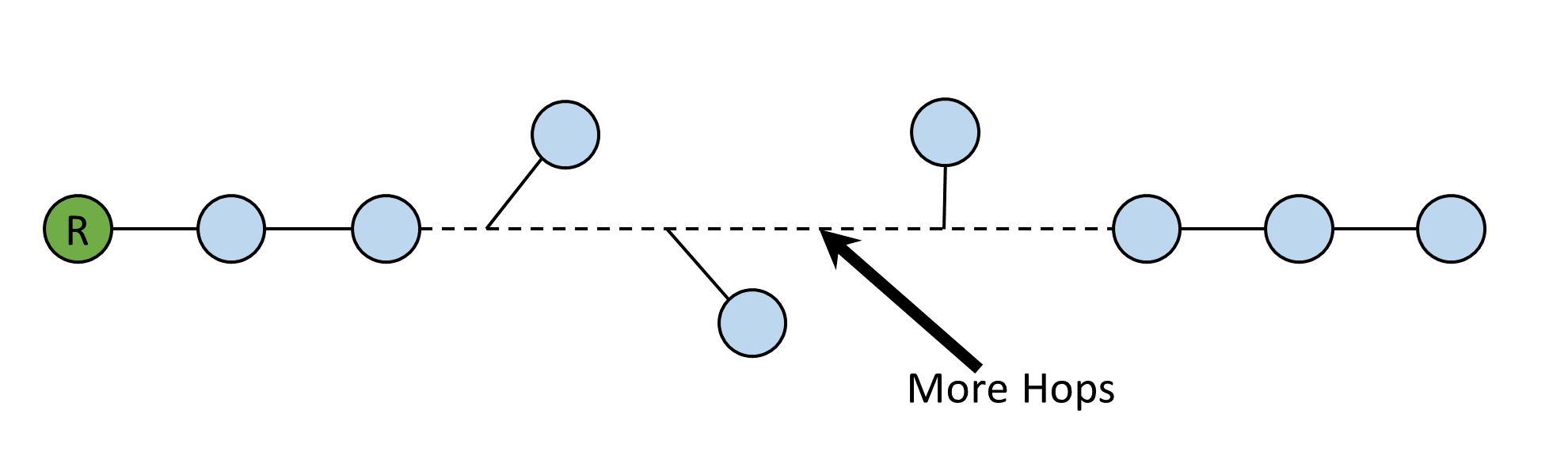}
	\caption{The illustration of the linear topology. Root is at the left end.}
	\label{fig:line}
\end{figure}

In the smart city scenario, urban structures may shape the network to a peculiar topology \cite{isrplready}. Inspired by \cite{isrplready}, we evaluate the scalability limit of the protocols in a linear network topology with small twigs, which may be quite common in urban areas. The linear network consists of 74 nodes and builds up to 68 hops, with the sink/root being at one end (as illustrated in Fig.~\ref{fig:line}). We use the Unit Disk Graph Medium (UDGM) with exponential distance loss as radio model and a maximum link quality of 90\% to account for uniform random noise. A node sends upward data packets randomly with an average interval of 10 minutes. After 20 minutes of network initialization, the sink starts to send an actuation packet every 10 seconds to a randomly selected target node. Upward packet payload is 60 bytes and downward packet payload is 20 bytes. Table~\ref{tab:ram} lists the compiled RAM usage of the test application with each protocol under different network size configurations. OSR was configured with the same MTU as the default in TinyRPL (i.e., 112 bytes). Since RPL storing mode has scalability issues in terms of memory when the routing table size increases, the routing table size in ContikiRPL (storing) and TinyRPL is configured to be 50, which results in 9104 bytes and 8160 bytes of memory footprint for ContikiRPL (storing) and TinyRPL, respectively. Each simulation ran for 4 hours, in which the total of 1320 downward packets were sent out. 

%
%

\begin{table}
	\caption{Comparison of RAM sizes for TelosB platform}
	\label{tab:ram}
	\begin{minipage}{\columnwidth}	
		\begin{tabular}{>{\centering}p{0.07\textwidth} >{\centering}p{0.12\textwidth} >{\centering}p{0.19\textwidth} >{\centering}p{0.19\textwidth} >{\centering}p{0.07\textwidth} >{\centering\arraybackslash}p{0.11\textwidth}}	
			
			\toprule
			\multirow{2}{1cm}{\bfseries WSN Size} & \multicolumn{5}{c}{\bfseries RAM Usage (bytes)} \\
			\cmidrule{2-6}
							&\bfseries  TinyRPL &\bfseries  ContikiRPL (S) &\bfseries  ContikiRPL (NS) & \bfseries ORPL &\bfseries  OSR \& CTP \\

			\midrule
			\bfseries 2	& 5952 & 7280 & \multirow{5}{*}{7164} 
							& \multirow{5}{*}{9710\footnote{ORPL includes a whole set of tools for logging.}} 
							& \multirow{5}{*}{3958} \\
			\bfseries 50 & 8160 & 9104 & \\
			\bfseries 74 & 9264 & 10016 & \\
			\bfseries 225 & +5972\footnote{Numbers with "+" indicates the amount that 
							overflowed the TelosB RAM space.} & +5516 &  \\
			\bfseries 400 & +14022 & +12164 &  \\
			
			\bottomrule
		\end{tabular}
	\end{minipage}
\end{table}

The evaluation results are shown in Table~\ref{tab:line}. OSR has successfully reached all the nodes (i.e., up to 68 hops) along the linear topology with 99.86\% downward PDR. In contrast, all RPL implementations suffer scalability problems. ContikiRPL (non-storing) only reached as far as 32 hops from the sink, far less than the theoretical threshold of 64 hops. Consequently, ContikiRPL (non-storing) has a poor PDR since more than half of the nodes are unreachable due to its scalability issue. On the other hand, the maximum reachable hop count in both ContikiRPL and TinyRPL storing modes as well as in ORPL is ad hoc, depending on the dynamics of nodes' limited routing table establishment. As we can see, their PDR performances are also significantly lower than that of OSR.  

\begin{table}
	\caption{Scalability Comparison on Linear Network}
	\label{tab:line}
	\begin{tabular}{ccc}
		\toprule
		\bfseries Protocol & \bfseries Max Reachable Hops & \bfseries PDR(\%) \\
		\midrule
		\bfseries OSR	& 68 & 99.86 \\
		\bfseries ContikiRPL (NS) & 32 & 33.31 \\
		\bfseries ContikiRPL (S) & -- & 76.91 \\
		\bfseries TinyRPL & -- & 27.95\\
		\bfseries ORPL & -- & 63.06 \\ 
		\bottomrule
	\end{tabular}
\end{table}

\begin{figure}
	\includegraphics{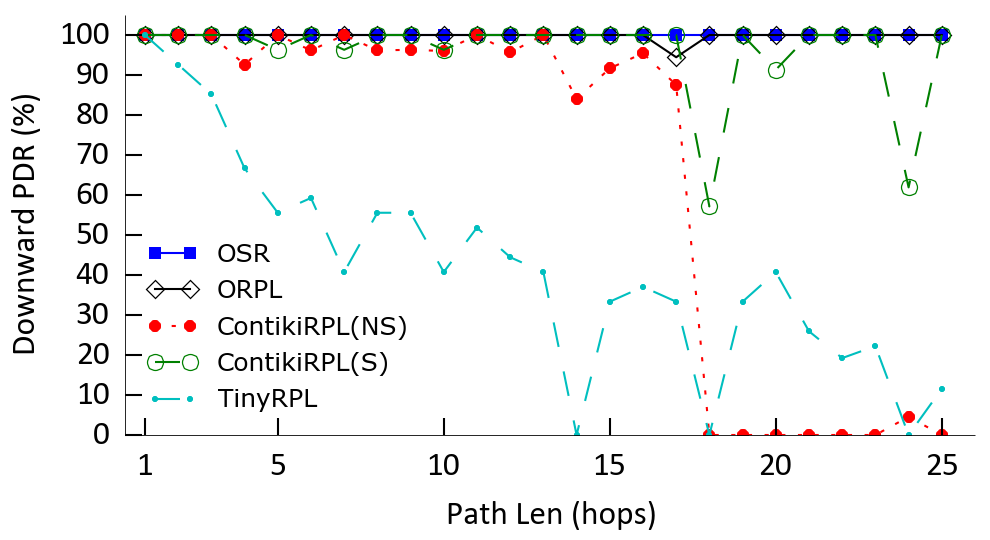}
	\caption{Downward PDR for linear topology based on downward path length in the first 25 hops.}
	\label{fig:hop_prr}
\end{figure}

To better understand about the protocols' scalability, we show in Fig.~\ref{fig:hop_prr} the network PDR up to the first 25 hops in the downward routing. As we can see, the PDR of TinyRPL drops quickly as the path length increases. ContikiRPL (non-storing) experiences a steep drop after the 17th hop, where the downward packet fragmentation begins due to the long downward path length. ContikiRPL (storing) maintains high PDR for most of the time, but suffers from performance drop at several random hops, which is likely due to the probabilistic occupation in their ancestor nodes' routing tables at times, which is also observed for TinyRPL. In contrast, OSR achieved nearly 100\% PDR regardless of the downward path length. ORPL also achieved high PDR within the first 25 hops. 

To summarize, ContikiRPL (non-storing) suffers scalability problem regarding the network diameter, whereas ContikiRPL (storing) and TinyRPL suffers scalability problem regarding the network size. IP fragmentation harms the performance of ContikiRPL (non-storing) significantly. ORPL also severely suffers from the scalability. We speculate that the network linear topology might have affected the anycast mechanism of ORPL. In contrast, OSR scales significantly better than all of the RPL implementations and ORPL. In fact, since OSR uses localized direct child set, it does not suffer as network size increases. Moreover, due to its Bloom filter based path encoding, OSR should be able to work with any path length of hundreds of hops.

\subsection{Evaluation in Indriya}
Next we evaluate the reliability and energy efficiency of the OSR in comparison with TinyRPL and Drip protocols in the Indriya testbed. ContikiRPL and ORPL were not included since they are based on the Contiki MAC and the datalink layer on Contiki platform which is very different from the TinyOS platform. As we know, energy efficiency is heavily dependent on the platform in addition to routing protocol\footnote{However, we have conducted simulations in Cooja to compare the relative energy efficiency of the protocols with their collection-only baseline. The test application is similar to that of Section 3.2, with a random topology. Compared to ContikiRPL baseline, the storing-mode has increased the node average duty cycle for 13.80\%, non-storing mode has increased the average node duty cycle for 10.34\%. Compared to a CTP baseline (in TinyOS), OSR has increased the average node duty cycle for 12.59\%.}. 

The Indriya testbed consisted of 95 TelosB nodes during the experiment time. The testbed was configured to be low power for our experiments. Node 31 at the corner on the first floor was selected as the sink to maximize the network diameter. The \textit{MAX\_BFLT\_LEN} was configured to 16 bytes. 

\subsubsection{Comparison against RPL}
Since TinyRPL could not work on the entire Indriya testbed, we conducted several experiment trials (30 minutes each) only using a half size of the testbed (i.e., 47 nodes with odd IDs) to evaluate OSR versus RPL. The test application collected data packets for the first 10 minutes in each trial for the network's initialization. The sink then sent downward packets to a randomly selected individual node every 10 seconds. Nodes were configured to be low power with a sleep interval of 1 seconds using the default TinyOS MAC (i.e., BoX-MAC \cite{boxmac}), whereas the sink was configured to be always on. We also conducted a pure CTP application as the baseline, where sink sends no downward packets, and node stops sending upward packets after the network's initialization.

Table~\ref{tab:OsrVsRpl} shows the performance results of OSR versus TinyRPL averaged on four trials. As we can see, OSR (with CTP) performs significantly better than TinyRPL on both the downward PDR and the duty cycle. TinyRPL's high duty cycle is mainly caused by its high DAO packet rate. We believe a careful tuning of the DAO rate could benefit TinyRPL's performance, however, it requires a systematic adjustment and is not the focus of this work. In particular, OSR itself only adds a very little to the duty cycle compared with the CTP baseline, demonstrating its energy efficiency. 

\begin{table}
	\caption{Comparison between OSR and TinyRPL on Low Power Indriya Testbed with 47 nodes}
	\label{tab:OsrVsRpl}
	\begin{tabular}{ccc}
		\toprule
		  &\bfseries PDR (\%) & \bfseries Duty Cycle (\%)\\
		\midrule
		\bfseries CTP Baseline	& --             & $4.26\pm0.07$ \\
		\bfseries TinyRPL         & $89.45 \pm 0.04$ & $18.67 \pm 0.45$\\
		\bfseries OSR            & $97.80 \pm 0.01$ & $4.43 \pm 0.17$ \\ 
		\bottomrule
	\end{tabular}
\end{table}

\subsubsection{Comparison against Drip}
Next, we compare OSR versus Drip in Indriya with all 95 available nodes. Drip is built on top of Trickle \cite{trickle} for dissemination the entire network. For unicast actuation, a \textit{target\_id} field is included in Drip's application packet to ensure only the targeted node would act when the command is received.

The WSN test application has a collection cycle interval of 4 minutes. Nodes operated on low power with a sleep interval of 1 second. The sink was configured to be always on. The sink started to issue one downstream packet to a randomly selected node every minute starting from the beginning of the third cycle if the node's upward packet was collected. Each experiment ran for about 6.8 hours, with 400 downward actuation packets were issued. 

Table~\ref{tab:OsrVsDrip} lists the downward performance comparison between OSR and Drip. Drip achieved 100\% downward PDR due to its flooding nature. OSR, on the other hand, achieved a PDR of 97.50\%. For energy efficiency, OSR (with CTP) achieved an average node duty cycle of 2.78\%, which is 45.81\% lower than the 5.13\% of Drip (with CTP). The result demonstrates that OSR is reliable and much more energy efficient than Drip. 

\begin{table}
	\caption{Comparison between OSR and Drip on Low Power Indriya Testbed with 95 nodes}
	\label{tab:OsrVsDrip}
	\begin{tabular}{ccc}
		\toprule
		& \bfseries PDR (\%) & \bfseries Duty Cycle (\%)\\
		\midrule
		\bfseries Drip        & 100   & $5.13 \pm 1.19$\\
		\bfseries OSR         & 97.50 & $2.78 \pm 1.26$ \\ 
		\bottomrule
	\end{tabular}
\end{table}

\subsection{Evaluation in TOSSIM}
We further conducted more simulations using TOSSIM \cite{tossim} for much larger network sizes and higher dynamics. We observed during the experiments in Indriya that the parent set size was one for most nodes when using CTP as the underlying collection protocol, since the communication environment of the indoor testbed is not very dynamic. As a result, the opportunistic routing of OSR could not be sufficiently evaluated when unicast failure occurs since there were few potential helper forwarders. 

We generated two networks of sizes 225 and 400 nodes uniformly distributed in a square area with the sink at a corner. The 400-node network was expanded from the 225-node network on both dimensions, retaining the same node density. The resulting network diameter was 11.58 hops and 19.07 hops, respectively. The collection cycle interval was 10 minutes. The sink sent a command packet to a randomly picked node every minute starting from the third cycle in simulation. The simulation terminated when 600 downstream packets were sent. The \textit{MAX\_BFLT\_LEN} was configured to 16 bytes and 20 bytes for 225-node and 400-node simulations, respectively. Since TOSSIM is targeted for MicaZ platform, the MTU was configured to 72 bytes to fit in MicaZ RAM space.

Most nodes in the simulations switched their parent nodes rapidly. On average, most nodes have more than 3 parents in their parent set with in a time window of 4 collection cycles. 

The evaluation results are shown in Table~\ref{tab:tossim}. Both tests achieved PDR above 98\%, which was not affected by the expansion of the network. The link unicast retransmission ratio for both tests were around 50\%, indicating a noisy and dynamic network condition. Regarding opportunistic routing, 3.33\% and 4.33\% of the packets experienced at least one active opportunistic routing occurrences (due to unicast delivery failure) in the 225-node simulation and 400-node simulation, respectively. On the other hand, both tests resulted in much more passive opportunistic routing occurrences (due to multiple matched children) than the active ones. We observed that for 225-node network, 13.33\% of the downward packets have experienced at least one instance of passive opportunistic routing, whereas for the 400-node network it was 36.33\%, about 3 times of that compared to the 225-node test, due likely to the larger network size and the longer downward path length. The occurrences of the opportunistic routing introduced about 15\% to 18\% duplicate traffic compared to the traditional source routing, which is inevitable due to the probabilistic nature of the Bloom filter (e.g., 9\% to 50\% duplicate traffic as reported in ORPL figure 6d \cite{orpl}).

\begin{table}[t]
\caption{OSR Performance in TOSSIM Simulation}
\label{tab:tossim}
\begin{minipage}{\columnwidth}
	\begin{tabularx}{\textwidth}{@{}YYYYYYY@{}}
		\toprule
		\multirow{2}{1cm}{\bfseries WSN Size} 
			& \multirow{2}{1cm}{\bfseries PDR (\%)} 
			& \multirow{2}{1cm}{\bfseries Max. Bflt (bytes)}
			& \multirow{2}{1cm}{\bfseries Ucast Retx} 
			& \multirow{2}{1.2cm}{\bfseries Dup. Traffic\footnote{With regard to link layer transmissions.} (\%)} 
			& \begin{tabular}[c]{c} \bfseries Pkt. with Opp. \\ \bfseries Routing \end{tabular} \\ 
		\cmidrule(l){6-7}
		 & & & & & \bfseries Active & \bfseries Passive  \\
		\midrule
		\bfseries 225 & 98.67 & 16 & 49.15\% & 15.32 & 3.33\%	& 13.33\%  \\ 
		\bfseries 400 & 98.67 & 20 & 51.69\% & 17.92 & 4.33\% & 36.33\% \\
		\bfseries 400 & 98.50 & 10 & 58.46\% & 37.31 & 4.83\% & 45.17\% \\
		
		\bottomrule
	\end{tabularx}
\end{minipage}
\end{table}

To test the effect of the maximum Bloom filter length, we also carried out simulation on the 400-node network with \textit{MAX\_BFLT\_LEN} being 10 bytes. The result is also shown in Table~\ref{tab:tossim}. As we can see, OSR also achieves the similar high PDR with a much smaller maximum Bloom filter length. However, due to a higher false positive rate, the duplicate traffic has increased to 37.31\%, which then caused more collisions, as indicated by the higher link unicast retransmission rate. Both the active and passive opportunistic routing occurrences have increased as well. 

%
%

\section{Discussion}
OSR has achieved reliable delivery of the downward unicast packets and desirable scalability as the network diameter and network size increase. OSR (with CTP) has also shown to achieve better energy efficiency compared to TinyRPL and Drip (with CTP) implemented on the same TinyOS platform. In this section we discuss our insights, and the limitations of current OSR implementation. 

\paragraph{Scalability}
Through our comprehensive evaluations, OSR achieves significantly better scalability compared to RPL storing and non-storing modes on two implementations and ORPL. OSR enables a very small and localized routing table compared to RPL storing mode and ORPL; simultaneously, OSR compresses the source-route effectively with respect to RPL non-storing mode. Therefore, OSR provides desirable scalability for resource-constrained real-world WSN/IoT deployments. 

\paragraph{Opportunistic routing}
OSR depends on the upward traffic to build the child/parent set. Thus, if the collection protocol is the best-path oriented, OSR may not be able to offer significant opportunistic routing due to the lack of potential helper forwarders in a relatively static communication environment. Even though in such a situation, OSR would degrade back to the traditional source routing in routing perspective, its compression of source-route in packet header is still exactly effective. OSR could benefit by working with a load balanced data collection protocol (e.g., CTP+EER \cite{eer}), which actively switches parents to balance the traffic load hence expands the parent set, in the case of static network condition. On the other hand, opportunistic routing in OSR introduces duplicate traffic to some degree due to its probabilistic nature. 

\paragraph{Link Asymmetry}
To build any downward path, OSR relies on upward routing tomography, whereas RPL non-storing mode uses the DAO messages to maintain the network topology at the root. Both approaches may lead to suboptimal downward path selection due to the link asymmetry. 

\paragraph{Interaction with IP}
We implemented OSR protocol, working with CTP in TinyOS, to validate our OSR approach for downward routing scalability and reliability. Thus, our current implementation lacks the ability to interact with IP like that of RPL and its variations. However, the principles of OSR can be readily applied to RPL non-storing mode to extend and improve its capability, which is considered in our future work. 

\paragraph{Differences from ORPL/CBFR}
OSR uses the Bloom filter to encode the source-route, whereas both ORPL and CBFR uses Bloom filter to compress the subgraph at each intermediate node. Each intermediate node stores its direct child set in OSR versus the subgraph of descendants in ORPL/CBFR. Also, OSR constructs the direct child set easily by intercepting the data collection packets, which is strictly localized and has negligible overhead. CBFR also intercepts the data collection packets, however, like ORPL, it requires the nodes in the network to exchange their Bloom filters to gather the subgraph information, introducing additional transmission overhead. Moreover, OSR can, to some extent, turn false positive into an opportunity in its passive opportunistic routing via local multicast. In CBFR, transmissions are broadcast based to explore all the possible downward paths to reach the destination node, resulting in a high transmission cost. On the other extreme, ORPL opportunistically selects only one node at a time for downward routing in the situation of multiple matches, hence it risks of delivering the packet to a false positive target node. An additional false positive recovery scheme is required in ORPL to address this problem, with increased delay. In contrast, OSR either opportunistically multicasts to all matched nodes, or opportunistically broadcasts when the downward link is broken. Hence, OSR is much less aggressive than CBFR, and has no need for false positive recovery in comparison with ORPL. The packet \textit{TTL} and duplicate suppression in OSR ensure that the packet forwarding by false positives would not chain forever. 

\section{Related Work}
Downward actuation protocols in WSNs can be classified into two categories: broadcast based and unicast/multicast based. A large portion of downward protocols, such as Drip \cite{drip}, Glossy \cite{glossy}, and Opportunistic Flooding \cite{oppflooding} are broadcast based that disseminate small data to the entire network (e.g., control packets). It would be very inefficient when using such broadcast based downstream protocols for individual node(s) actuation in LLNs, since the actuation commands would have to be flooded over the entire network. Recently, the demand to individual node(s) actuation arises as heterogeneity becomes popular in WSN/IoT deployments, in which individual nodes play different roles in the network. RoCoCo \cite{rococo} integrates the data collection and command dissemination. The command information is piggybacked in the routing beacons of the collection protocol, where the receivers' addresses are also included, hence enables dissemination to a subset of nodes. CBFR \cite{cbfr} and RBD \cite{rbd} utilize the tree structure for downward routing. At each intermediate node, the route is determined by checking the whole subtree information stored at that node. Whereas RBD stores the raw addresses of the nodes in the subtree, CBFR utilizes counting Bloom filter to reduce the memory overhead and supports gradual forgetting for nodes mobility.  

RPL \cite{rplrfc} is a recent standard routing protocol for LLNs that supports upward, downward, and point-to-point (P2P) traffic patterns. RPL supports two modes for downward traffic, either through the entire subtree stored at each intermediate node's routing table (storing mode), or through the source-route specified at the root (non-storing mode). Storing mode suffers scalability problem with regard to the network size due to the limited memory in resource-constrained nodes, whereas non-storing mode suffers scalability problem with regard to the network diameter, due to the limitation in the frame size of the LLNs. A recent and comprehensive survey of RPL can be found in \cite{rplsurvey17}. ORPL \cite{orpl} brings opportunistic routing into RPL and improves the performance of RPL. Similar to CBFR, ORPL adopts Bloom filter/bitmap to represent node's subgraph (i.e., routing table) to reduce the memory overhead. ORPL using Bitmap only works for predefined static networks. When using ORPL on dynamic networks, the Bloom filter compression of a node's subgraph is propagated upward the collection tree in order for parents to update their routing tables. When the network is large, the Bloom filter size may also grow quickly and it would be inefficient for nodes to exchange their Bloom filters. Hence ORPL and CBFR also suffer from the scalability problem for large-size WSNs/IoT.

HB-DSR \cite{hashdsr} is known to be the first source routing protocol that encodes the route into a Bloom filter. Bloom filters have been used in several approaches for multicast, such as \cite{fwdAnomlies} and \cite{stageBflt}. However, those approaches are only targeted for wireline networks or mobile networks without significant resource constraints. 

After our OSR development, we learned that IETF Roll Working Group is working on a draft Constrained-Cast \cite{ccast} to consider using the Bloom filter to encode the source-routes in RPL (non-storing mode) for forwarding multicast traffic. However, this draft is still in the process and many details are unclear. Besides, this draft defines a few possible values of the Bloom filter size, as opposed to our adaptive Bloom filter size in OSR. 

\section{Conclusion and Future Work}
By introducing opportunistic routing into the traditional souring routing approach, we presented OSR, an opportunistic-supported source routing approach and protocol which provides reliable and scalable downward actuation in large-scale WSN/IoT systems. The unique opportunistic nature of OSR effectively addresses the fundamental issues of the drastic wireless link dynamics in noisy and resource-constrained WSNs. OSR only stores the direct child set at each intermediate node rather than the entire subtree of descendants as other address-based routing protocols. As a result, OSR has small memory overhead and achieves great scalability while maintaining good performance. The results on our simulations and real-world WSN testbed experiments demonstrate the merits of OSR. OSR significantly outperforms RPL storing mode and non-storing mode on two most widely used implementations. On the other hand, while OSR achieves desirable and comparable packet delivery rate as the flooding based Drip, it has much lower duty cycle in comparison with Drip. Our future work includes to extend OSR for WSN downward multicast routing, and to apply/integrate OSR with RPL non-storing mode. We believe OSR provides a significant and practical solution to wireless actuation for large-scale and resource-constrained WSN/IoT deployments.

\section*{Acknowledgments}
This work is supported in part by U.S. National Science Foundation under grant CNS-1320132.

\bibliographystyle{acm}


\bibliography{osr_latex} 

\end{document}